\documentclass[prd,aps,superscriptaddress,amsmath,amssymb,nobibnotes,nofootinbib,10pt]{revtex4-2}

\pdfoutput=1 

\usepackage[T1]{fontenc} 
\usepackage{amsmath,amssymb,mathtools}
\usepackage{color}
\usepackage{hyperref}

\hypersetup{
    colorlinks=true,
    linkcolor=blueG,
    citecolor=redG,
    urlcolor=magentaG,
    bookmarks=true
}
\usepackage{graphicx}
\usepackage{xcolor}
\usepackage{braket}
\usepackage{comment}
\usepackage{tikz}

\definecolor{blueG}{RGB}{51, 102, 204}
\definecolor{magentaG}{RGB}{214, 40, 132}
\definecolor{redG}{RGB}{229, 51, 102}

\newcommand{\nc}{\newcommand}
\nc{\ir}{\mathrm{i}}
\nc{\dd}{\mathrm{d}} 
\nc{\eE}{\mathsf{e}}
\nc{\Tr}{\text{Tr}}
\nc{\tdet}{\tilde{\det}}
\nc{\J}{\mathcal{J}}
\nc{\B}{\mathcal{B}}
\nc{\N}{\mathcal{N}}
\nc{\K}{\mathcal{K}}
\nc{\st}{\omega}
\nc{\Gammax}{\Gamma_{\mathrm{x}}}
\nc{\purple}[1]{\textcolor{purple}{#1}}

\usepackage[normalem]{ulem}

\begin{document}

\title{Entanglement of Inhomogeneous Free Bosons and Orthogonal Polynomials}

\author{Pierre-Antoine Bernard}
\affiliation{Centre de Recherches Math\'ematiques, Universit\'e de Montr\'eal, P.O. Box 6128, Centre-ville Station, Montr\'eal (Qu\'ebec), H3C 3J7, Canada}

\author{Rafael I. Nepomechie}
\affiliation{Department of Physics, P.O. Box 248046, University of Miami, Coral Gables, FL 33124 USA}

\author{Gilles Parez}
\email{parez@lapth.cnrs.fr}
\affiliation{Laboratoire d'Annecy-le-Vieux de Physique Th\'eorique (LAPTh), CNRS, Universit\'e Savoie Mont Blanc, 74940 Annecy, France}

\author{\'Eric Ragoucy}
\affiliation{Laboratoire d'Annecy-le-Vieux de Physique Th\'eorique (LAPTh), CNRS, Universit\'e Savoie Mont Blanc, 74940 Annecy, France}

\author{David Raveh}
\affiliation{Department of Physics and Astronomy, Rutgers University, Piscataway, NJ 08854-8019 USA}

\author{Luc Vinet}
\affiliation{Centre de Recherches Math\'ematiques, Universit\'e de Montr\'eal, P.O. Box 6128, Centre-ville Station, Montr\'eal (Qu\'ebec), H3C 3J7, Canada}
\affiliation{D\'epartement de Physique, Universit\'e de Montr\'eal, Montr\'eal (Qu\'ebec), H3C 3J7, Canada}
\affiliation{IVADO, 6666 Rue Saint-Urbain, Montr\'eal (Qu\'ebec), H2S 3H1, Canada}

\begin{abstract}
In this paper, we investigate the ground-state entanglement entropy in inhomogeneous free-boson models in one spatial dimension. We develop a powerful method to extract the leading term in the entanglement scaling, based on the analytic properties of the inhomogeneous potential. This method is applicable to a broad class of models with smooth spatial inhomogeneities. As a case study, we apply this approach for a family of exactly-solvable models characterized by orthogonal polynomials of the Askey scheme, finding a perfect match between the numerical and analytical results.
\end{abstract}

\maketitle

\tableofcontents

\section{Introduction}

Entanglement is a fundamentally quantum phenomenon, with no direct classical analogue. Over the past few decades, it has emerged as a central theme across a wide range of research fields, including quantum information, statistical mechanics, condensed matter physics, and high energy physics. In particular, the study of entanglement in quantum many-body systems has provided critical insights into complex emergent phenomena \cite{amico2008entanglement,Laflorencie:2015eck}, such as quantum phase transitions \cite{OAFF02,VLRK03}, topological phases of matter \cite{kitaev2006topological,2006PhRvL..96k0405L}, and non-equilibrium dynamics \cite{cc-05,ac-17,c-20}.

Numerous quantitative measures of entanglement exist, depending on the physical context and the choice of entangled subsystems. For a quantum systems in a pure state $|\psi \rangle$, the entanglement between a subsystem $A$ and its complement $B=\bar A$ is typically quantified by the entanglement entropy, 
\begin{equation}
    S_A = -\Tr (\rho_A \log \rho_A), \qquad \rho_A = \Tr_B |\psi \rangle \langle \psi|, 
\end{equation}
where $\rho_A$ is the reduced density matrix of system $A$. This quantity has been the focus of extensive research across diverse physical contexts, with particular emphasis on ground states of quantum many-body Hamiltonians. For gapped Hamiltonians in arbitrary dimensions, the ground-state entanglement entropy obeys an area law \cite{hastings2007area, wolf2008area, eisert2010colloquium}.
In contrast, the situation is more intricate for gapless systems. In one-dimensional quantum critical systems with an underlying conformal field theory (CFT) \cite{francesco2012conformal}, the entanglement entropy diverges logarithmically with the subsystem size and is proportional to the central charge $c$ of the CFT. Assuming that $A$ is a contiguous segment of length $\ell$ embedded in an infinite or semi-infinite system, we have the celebrated Calabrese-Cardy formula \cite{CC04,Calabrese:2009qy}
\begin{equation}
    S_A \sim \frac{r c}{6} \log \ell
\end{equation}
for large $\ell$, where $r$ is the number of contact points between $A$ and the rest of the system. We have $r=2$ when $A$ is embedded in an infinite system, whereas $r=1$ if $A$ and the complementary are both semi-infinite and are adjacent in one point. 

For finite critical lattice systems, the entropy depends on the subsystem size $\ell$ and on the total system size $N+1$. 
In the case where $A$ is an interval attached to a boundary of a homogeneous system, we have \cite{CC04,Calabrese:2009qy}
\begin{equation}\label{eq:SAScaling}
    S_A = \frac{c}{6}\log \left(\frac{2(N+1)}{\pi}\sin\Big(\frac{\pi \ell}{N+1}\Big)\right) + s_0
\end{equation}
at leading order in the large-$N$ limit, where $s_0$ is a non-universal constant. For certain inhomogeneous models, the entanglement scaling can be understood in the framework of curved-space CFT \cite{DSVC17}, where the various lengths in Eq.~\eqref{eq:SAScaling} are modified by the space curvature.

Since these pioneering results, exactly solvable lattice models have emerged as a rich playground for exploring further the nature of quantum entanglement in many-body system. Free fermionic and bosonic models, with their associated analytical methods \cite{peschel2003calculation,peschel2009reduced, casini2009entanglement}, have been used to probe entanglement structure in various new contexts, including systems with interface and defects~\cite{peschel2005entanglement}, quantum quenches \cite{fc-08,parez2022analytical}, systems in higher dimensions \cite{berthiere2019boundary,mrc-20,bernard2022entanglement,parez2024entanglement}, symmetry-resolved entanglement~\cite{bons,murciano2020symmetry,PBC21}, entanglement Hamiltonians \cite{EP-17,di2020entanglement,BE-24}, mixed-states entanglement \cite{de2016entanglement,SSR17,SRRC19,SR19,ac2-22,angel2020logarithmic,berthiere2023reflected,berthiere2023reflectedB} and systems with spatial inhomogeneities \cite{eisler2009entanglement,Ramirez:2015yfa,Rodriguez-Laguna:2016roi,Crampe:2019upj,Finkel:2020lgf,Finkel:2021gji}. For the latter, in the case of free fermions, an approach based on orthogonal polynomials has been proposed in \cite{Crampe:2019upj,Crampe:2021}. There, the spatial inhomogeneities of the model are fine-tuned to match the recurrence relations of orthogonal polynomials of the Askey scheme \cite{Koekoek}. It allows for their exact diagonalization and the calculation of various entanglement measures, matching curved-CFT calculations \cite{bernard2022entanglement,bernard2024entanglement,blanchet24Neg,bernard2024distinctive,bernard2024entanglementB}.

Inhomogeneous bosonic models have also been investigated \cite{dubail2017emergence,brun2017one,brun2018inhomogeneous}, although comparatively less extensively than their fermionic counterparts. 
The goal of this paper is to reduce this gap, and provide strong tools to understand the entanglement structure in arbitrary inhomogeneous bosonic systems. In particular, we   
propose a simple method for computing the leading behavior of the entanglement entropy for large $N$,
based on the notion of the effective size $\N_{\rm eff}$ of the inhomogeneous system, directly related to analytical properties of the inhomogeneous potential. 
To illustrate these general results, we introduce a family of exactly-solvable free-boson models based on orthogonal polynomials of the Askey scheme, and study the scaling of the entanglement entropy both analytically and numerically, finding a perfect match between the two approaches. 

This paper is organized as follows. In Sec.~\ref{sec:hom} we review the homogeneous free-boson harmonic chain with open boundary conditions, and the methods to compute the ground-state entanglement from two-point correlation functions. As a warm-up, we compute the entanglement entropy both in the massive and massless regime and observe the expected scaling from CFT. Going beyond the homogeneous case, we introduce a generic inhomogeneous free-boson model in Sec.~\ref{sec:inhMod} and describe our general method to extract the scaling of the entanglement entropy from the analytical properties of the inhomogeneous potential. Section~\ref{sec:example} is devoted to concrete illustrations of this general method. We introduce exactly-solvable versions of the generic inhomogeneous model by fine-tuning the parameters. In particular, we obtain two models, related to Krawtchouk and dual Hahn polynomials, respectively. In both cases, we find a perfect agreement between the entanglement scaling predicted by our general method and the exact numerical results obtained from the analytical diagonalization of the models. Finally, we introduce a one-parameter model that interpolates between the Krawtchouk and dual Hahn cases. Here, our analytical method predicts a continuum of effective lengths, and hence of entanglement scaling coefficients, and we confirm those predictions with exact numerical results. We give an overview of the results and discuss opportunities for future research projects in Sec.~\ref{sec:ccl}. 

\section{Review of the homogeneous chain}\label{sec:hom}

In this section, we review the homogeneous free-boson chain, its definition, diagonalization and corresponding results for the entanglement entropy. 

\subsection{Definition and diagonalization}

We consider a one-dimensional chain of free-bosons with mass $m$ and open boundary conditions. By convention, we label the sites with $x=0,1,\dots,N$. The Hamiltonian reads
\begin{equation}
    H = \frac 12 \sum_{x=0}^N \pi_x^2 + \frac12 \sum_{x=0}^{N-1} (\phi_{x+1}-\phi_x)^2 +  \frac12 \sum_{x=0}^N m^2 \phi_x^2 \,,
\label{inhomogHam0}
\end{equation}
where the operators $\pi_i$ and $\phi_i$ are real and verify the canonical commutation relations
\begin{equation}
    [\phi_j, \pi_k]=\ir \delta_{jk}, \quad [\phi_j, \phi_k]= [\pi_j, \pi_k]=0.
\end{equation}

We recast the Hamiltonian using a tridiagonal Hermitian matrix $\mathcal{K}$,
\begin{equation}
    H = \frac 12 \sum_{x=0}^N \pi_x^2+\frac 12 \sum_{x,y=0}^N \phi_x \mathcal{K}_{xy} \phi_y \ ,
    \label{Hamiltonian0}
\end{equation}
where the associated matrix $ \mathcal{K}$ is given by
\begin{equation}
     \mathcal{K} = \begin{pmatrix}
           m^2+2 & -1       &         &        &       &    \\
           -1     &m^2+2    &-1       &        &       &   \\
           &-1       & m^2+2      &  -1        &       &         \\
            &       &\ddots   &\ddots   &\ddots & & \\
           &        &    &     -1       & m^2+2     &  -1 \\ 
           &        &   &            & -1  &m^2+2       
    \end{pmatrix} \,.
    \label{Khomog}
\end{equation}
 We write the matrix $ \mathcal{K}$ in diagonal form, $ \mathcal{K}=U \Lambda U^\dagger$, where $\Lambda_{kk'} = \delta_{kk'}\Lambda_k$ and $U$ is a unitary matrix with entries $U_{xk} = u_x(\Lambda_k).$ 
The diagonalization of this matrix is standard. The eigenvalues read
\begin{equation}
\Lambda_k=2+m^2-2\cos\left(\frac{(k+1)\pi}{N+2}\right)\,,
\quad k=0,1,\ldots,N \,,
\end{equation}
and corresponding normalized eigenvectors are
\begin{equation}
u_{x}\left(\Lambda_{k}\right)=\sqrt{\frac{2}{N+2}}\sin\left(\frac{(x+1)(k+1)\pi}{N+2}\right), 
\end{equation}
with $x,k=0,1,\ldots,N.$ These results can also be understood within the framework of discrete orthogonal polynomials, namely a certain discretization of the (continuous) Chebyshev polynomials of the second kind \cite{Chebyshev, Zhedanov:2019}, see Ref. \cite{Crampe:2019upj}. 

To obtain the spectrum of the Hamiltonian $H$, we introduce new sets of coordinates and momenta based on the eigenvectors of $\mathcal{K}$, 
\begin{equation}
    \varphi_k = \sum_{x=0}^N \phi_x u_x\left(\Lambda_k\right) ,\qquad
    \varpi_k =  \sum_{x=0}^N \pi_x u_x\left(\Lambda_k\right). 
\end{equation}
These new operators satisfy the canonical commutation relation, $[\varphi_k, \varpi_{k'}]=\ir\delta_{kk'}$. In these coordinates, the Hamiltonian describes $N+1$ decoupled oscillators, 
\begin{equation}
    H = \frac 12 \sum_{k=0}^N \varpi_k^2+\frac 12 \sum_{k=0}^N \Lambda_k \varphi_k^2.
\end{equation}

Next, defining $\lambda_k \equiv \sqrt{\Lambda_k}$, we introduce the creation and annihilation operators
\begin{equation}
    a_k = \frac{1}{\sqrt{2}}(\sqrt{\lambda_k}\varphi_k + \frac{\ir}{\sqrt{\lambda_k}}\varpi_k), \qquad
    a_k^\dagger = \frac{1}{\sqrt{2}}(\sqrt{\lambda_k}\varphi_k - \frac{\ir}{\sqrt{\lambda_k}}\varpi_k) \,,
    \label{aadag}
\end{equation}
and find
\begin{equation}\label{eq:Hdiag}
    H = \sum_{k=0}^N \lambda_k \Big(a_k^\dagger a_k+\frac 12\Big). 
\end{equation}
Importantly, the eigenvalues are positive, $\lambda_k > 0$. For a positive mass $m>0$, the eigenvalues are strictly positive and the model is gapped. For the massless case $m=0$, the lowest eigenvalues approach $0$ in the thermodynamic limit $N\to \infty$, and the model becomes gapless. Since all eigenvalues are positive, the ground state of the model is the vacuum $|0\rangle$. It satisfies $a_k|0\rangle = 0$ for all $k=0,1,\dots, N$.

\subsection{Entanglement entropies from correlation matrices}\label{sec:EECa}
In this section, we review how to express ground-state entanglement entropies in terms of two-point correlation functions, as discussed in Ref. \cite{casini2009entanglement}. First, we consider the following two ground-state correlation functions,
\begin{equation}
\begin{split}
    X_{ij} &= \langle 0|\phi_i \phi_j|0 \rangle, \\[.2cm]
     P_{ij} &= \langle 0| \pi_i \pi_j|0 \rangle. 
    \end{split}
\end{equation}
Importantly, they satisfy
\begin{equation}
\label{eq:corrXP}
\begin{split}
    X_{ij} &= \frac12 [\mathcal{K}^{-1/2}]_{ij} = \sum_{k=0}^N \frac{1}{2\lambda_k} U_{ik}U_{jk}  \,, \\
     P_{ij} &=  \frac12 [\mathcal{K}^{1/2}]_{ij} = \sum_{k=0}^N \frac{\lambda_k}{2} U_{ik}U_{jk} \,,
    \end{split}
\end{equation}
where the matrix $U$ is defined below Eq.~\eqref{Khomog}.

Second, we define the chopped correlation matrices $X_A,P_A$ as $[X_A]_{ij}=X_{ij}$ for $i,j \in A$, and similarly for~$P_A$. 
From $X_A,P_A$, one defines the matrix \cite{casini2009entanglement} 
\begin{equation}
    C_A=(X_A P_A)^{1/2}\,.
    \label{eq:corrC}
\end{equation}
One can show  \cite{casini2009entanglement} that $C_A\geqslant 1/2$, which ensures that the entanglement entropy, given by 
\begin{equation}
    S_A =  \Tr\Big[(C_A+1/2)\log (C_A+1/2)-(C_A-1/2)\log (C_A-1/2)\Big],
    \label{eq:SA}
\end{equation}
is well-defined.

\begin{figure}
    \centering
   \includegraphics[width=0.45\linewidth]{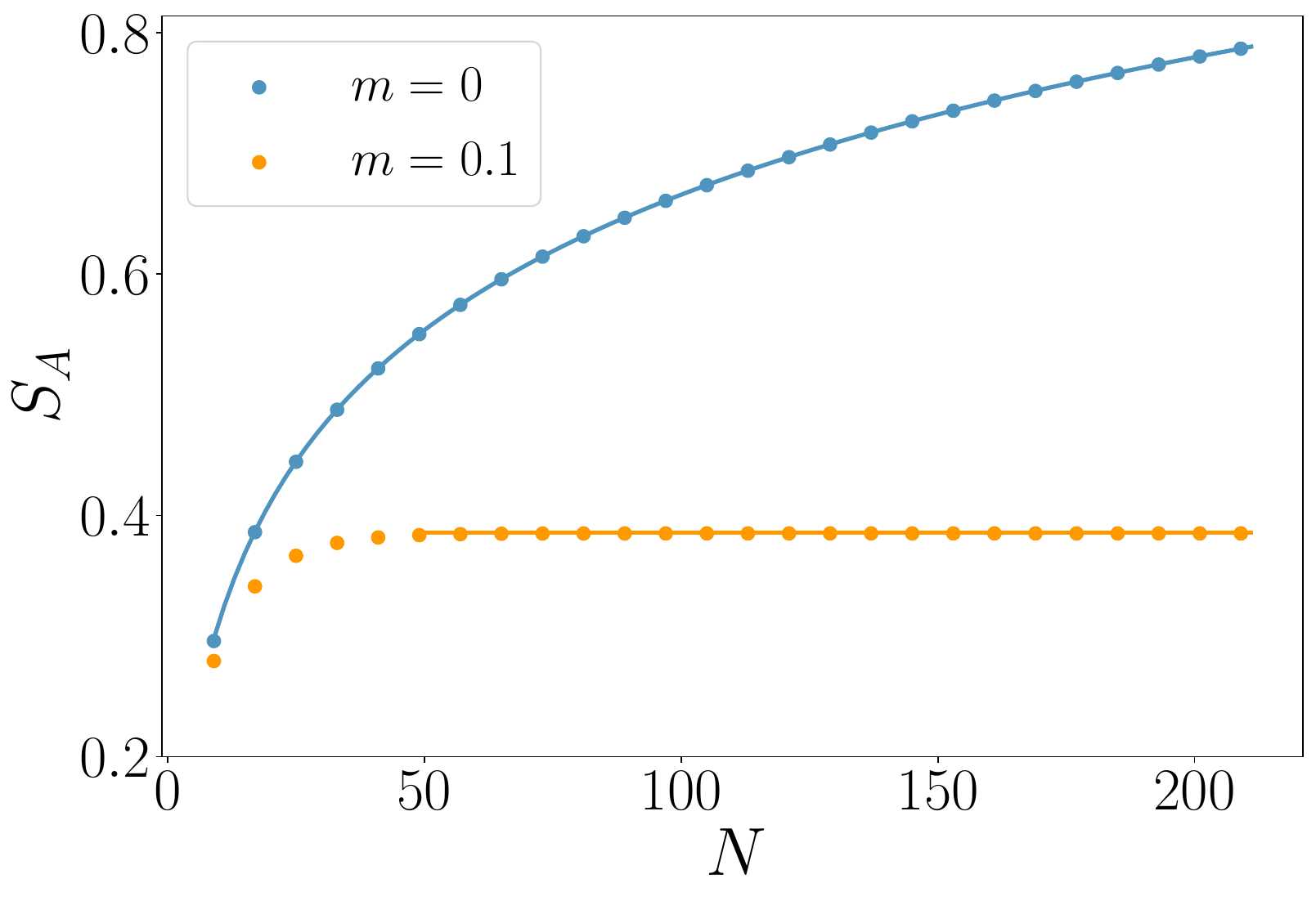}\hspace{.5cm}
   \includegraphics[width=0.45\linewidth]{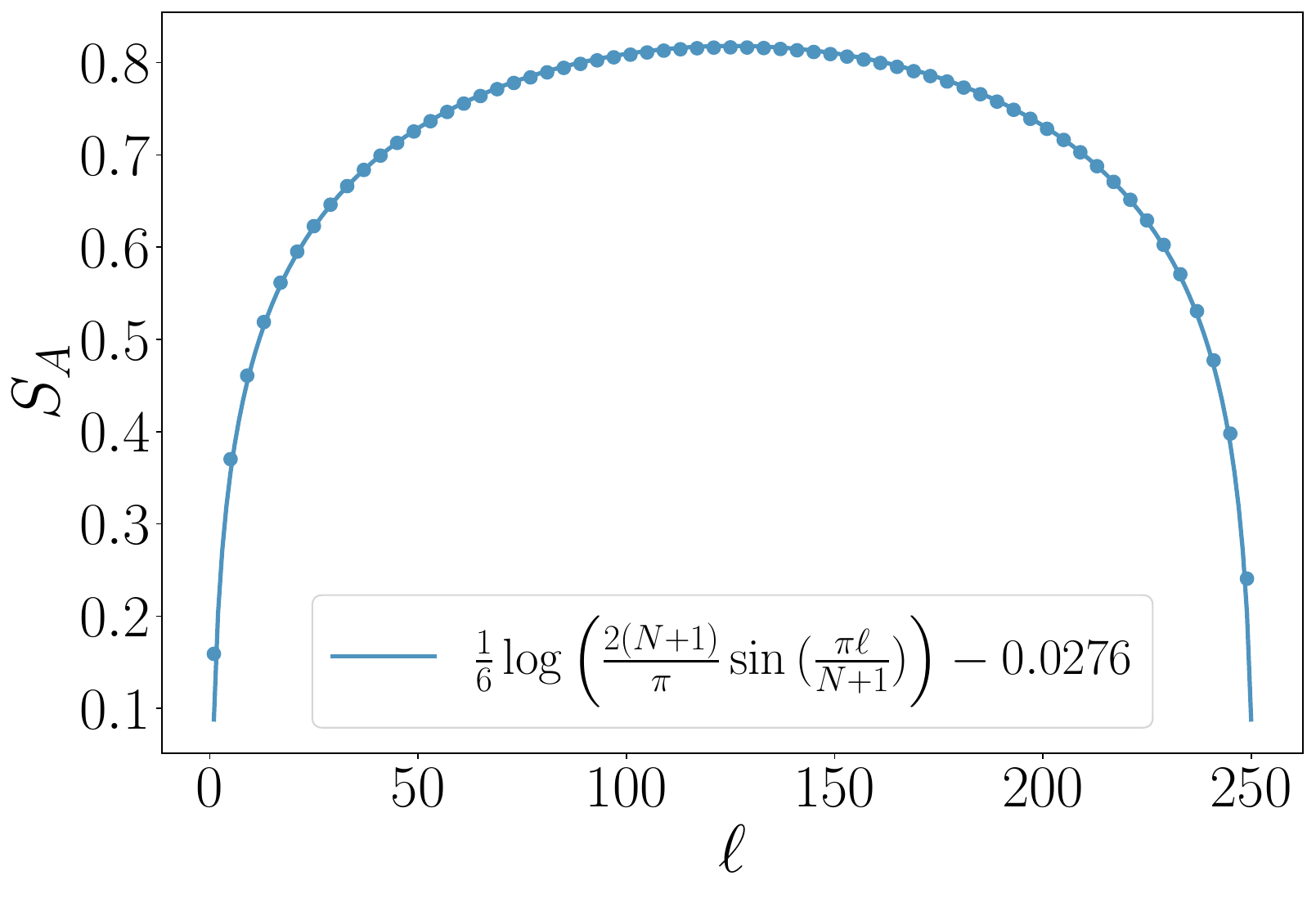}
    \caption{\textit{Left:} Entanglement entropy $S_A$ as a function of $N$ with subsystem size $\ell=N/2.$ In the gapless case (blue curve) the entropy diverges logarithmically with $N$ as $S_A \sim \frac 16 \log N$, whereas in the gapped case (orange curve) it is a constant in the large-$N$ limit. \textit{Right:} Entanglement entropy $S_A$ as a function of subsystem size $\ell$ with fixed $N=250$ in the massless case. The match between the numerical data (symbols) and the prediction of Eq.~\eqref{eq:SAScaling} with $c=1$ and  $s_0=-0.0276$ (solid line) is extremely convincing. }
    \label{fig:S1Hom}
\end{figure}

Using the results above, 
we investigate the ground-state entanglement entropy between two complementary segments of lengths $\ell$ and $N-\ell+1$, namely $A=\{0,1,\dots,\ell-1\}$ and $B=\{\ell,\ell+1,\dots,N\}$. 
In the massless case $m=0$, the model is gapless and is described by a CFT with central charge $c=1$. 
We find that the entanglement entropy scales as in Eq.~\eqref{eq:SAScaling} with $s_0=-0.0276$. For $m>0$, the theory is gaped and the entanglement entropy satisfies an area law, which is a constant in one dimension. We illustrate these results in Fig.~\ref{fig:S1Hom}. %

\section{Results for the inhomogeneous chain}\label{sec:inhMod}

In this section, we focus on free-boson chains with inhomogeneous couplings. 

\subsection{The model}
We consider a chain of $N+1$ coupled quantum oscillators with open boundary conditions, described by the Hamiltonian
\begin{equation}
    H = \frac 12 \sum_{x=0}^N \pi_x^2 + \frac12 \sum_{x=0}^{N-1} J_x (\phi_{x+1}-\phi_x)^2 +  \frac12 \sum_{x=0}^N V_x \phi_x^2 \,.
\label{inhomogHam}
\end{equation}
Here, $J_x,V_x\geqslant 0$ are the inhomogeneous coupling strength between the sites $x$ and $x+1$ and local potential at site~$x$, respectively. The mass term $m^2$ that appears in the homogeneous Hamiltonian \eqref{inhomogHam0} is here absorbed into the potential~$V_x$. Similarly as in the homogeneous case, we recast the Hamiltonian using a tridiagonal Hermitian matrix~$\mathcal{K}$,
\begin{equation}
    H = \frac 12 \sum_{x=0}^N \pi_x^2+\frac 12 \sum_{x,y=0}^N \phi_x  \mathcal{K}_{xy} \phi_y \ ,
    \label{Hamiltonian}
\end{equation}
where
\begin{equation}
     \mathcal{K} = \begin{pmatrix}
           B_0 & -J_0       &         &        &       &    \\
           -J_0     &B_1    &-J_1       &        &       &   \\
           &-J_1       & B_2      &  -J_2        &       &         \\
            &       &\ddots   &\ddots   &\ddots & & \\
           &        &    &     -J_{N-2}       & B_{N-1}     &  -J_{N-1} \\ 
           &        &   &            & -J_{N-1}  &B_N        
    \end{pmatrix} \, ,
    \label{Kinhomog}
\end{equation}
with 
\begin{equation}\label{eq:Vx}
     B_x\equiv V_x+J_x+J_{x-1}
\end{equation}
and we use the convention $J_{-1}=J_N=0$. 

In the following, we systematically choose the parameters $J_x,V_x$ such that $ \mathcal{K}$ is a positive-definite matrix, with strictly positive eigenvalues $\Lambda_k > 0, \ k=0,1,\dots, N.$  As in the homogeneous case, the matrix $ \mathcal{K}$ is cast in diagonal form $ \mathcal{K}=U \Lambda U^\dagger$, where $\Lambda_{kk'} = \delta_{kk'}\Lambda_k$ and $U$ is a unitary matrix with entries $U_{xk} = u_x(\Lambda_k).$ These entries satisfy the recurrence relation
\begin{equation}
-\Lambda_k u_x\left(\Lambda_k\right)= J_x u_{x+1}\left(\Lambda_k\right)-B_x u_x\left(\Lambda_k\right)+J_{x-1}u_{x-1}\left(\Lambda_k\right)\,
\label{urec}
\end{equation}
for $x=0,1,\dots, N,$ as well as the orthogonality condition 
\begin{equation}
\sum_{x=0}^N u_x\left(\Lambda_k\right)\, u_x\left(\Lambda_{k'}\right)=\delta_{kk'}\,.
\label{unorm}
\end{equation}

\subsection{Inhomogeneous potential}\label{sec:potential}

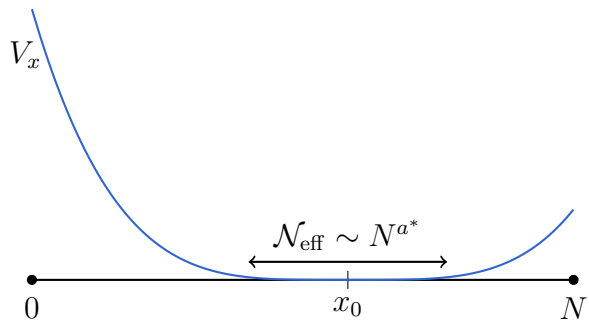
\begin{figure}
    \centering
    \begin{tikzpicture}[scale=1.2]

\draw[thick] (0,0) -- (6,0);
\draw (0.2,2.5) node[left] {\large $V_x$};

\node[below] at (0, -0.1) {\large $0$};
\draw[black, fill=black] (0,0) circle(0.05cm);
\draw (3.5,0.1) -- (3.5,-0.1) node[below] {\large $x_0$};
\node[below] at (6, -0.1) {\large $N$}; 
\draw[black, fill=black] (6,0) circle(0.05cm);

\draw (3.5,0.5) node{\large $\N_{\rm eff} \sim N^{a^*}$};
\draw[<->, thick] (2.4,0.2)--(4.6,0.2);

\draw[domain=0:6,samples=100,smooth,thick,blueG, thick] 
    plot (\x, {0.02*(\x-3.5)^4}); 

\end{tikzpicture}
\caption{Illustration of a generic smooth inhomogeneous potential $V_x$ vanishing at $x=x_0$ with an effective massless region of size $\N_{\rm eff}\sim N^{a^*}$.}
    \label{fig:VxGen}
\end{figure}

In the inhomogeneous model described in Eq.~\eqref{inhomogHam}, the potential $V_x$ can be interpreted as a position-dependent effective mass (squared). Therefore, regions where $V_x$ vanishes correspond effectively to a locally massless theory. Let us assume that the potential vanishes at $x=x_0$ in the large-$N$ limit. We investigate the potential in the vicinity of~$x_0$, and define $\mathcal I = [0, a^*)$ with $a^*\leqslant 1$ as the half-open interval such that we have
\begin{equation}
    \lim_{N\to \infty} V_{x_0 \pm y N^a} = 0 \qquad \forall a \in \mathcal I,
\end{equation}
where $y$ is an arbitrary real parameter satisfying $x_0 + y N^a\leqslant N$ and  $x_0 - y N^a\geqslant 0$. This indicates that there is a region of effective size $\N_{\rm eff} \sim N^{a^*}$ centered on $x_0$ where the theory is effectively massless in the thermodynamic limit. We illustrate this in Fig.~\ref{fig:VxGen}. We thus expect the entanglement entropy of the ground state between the subsystems $A=\{0,1,\dots, x_0\}$ and $B=\{x_0+1,\dots, N\}$, to scale as 
\begin{equation}
\begin{split}
    S_{A} &\sim \frac{1}{6} \log \N_{\rm eff} \\ & \sim \frac {a^*}6 \log N
    \end{split}
\end{equation}
in the large-$N$ limit. This corresponds to the scaling of the homogeneous case, multiplied by ~${a^*}$, which we define as the effective central charge, $c_{\rm eff}={a^*}$. 
The term \textit{effective central charge} is used here in analogy with its appearance in the context of entanglement with interface defect, as discussed in \cite{peschel2005entanglement}, and does not imply an underlying conformal invariance. In particular, we do not identify a conformal invariance in the models we study, apart from the homogeneous case. We note that the above results also hold when  $A=\{0,1,\dots, x\}$ and $B=\{x+1,\dots, N\}$ are adjacent at a point $x$ such that $|x-x_0|<\mathcal{O}(\N_{\rm eff})$, and hence $x$ lies in the effective massless region in the thermodynamic limit.

In contrast, when $|x-x_0|>\mathcal{O}(\N_{\rm eff})$, $A$ and $B$ are adjacent in a locally massive region, and we thus expect the entanglement entropy between such regions to scale as a constant, similarly to massive homogeneous bosons.

\section{Examples and orthogonal polynomials}\label{sec:example}

In this section, we illustrate the general predictions discussed above in different inhomogeneous models based on orthogonal polynomials of the Askey scheme.

\subsection{Orthogonal polynomials of the Askey scheme}

Orthogonal polynomials of Racah type form an important family within the classical discrete orthogonal polynomials. These polynomials were introduced 
in the context of angular momentum coupling in quantum mechanics~\cite{racah1942theory}. They are part of the Askey scheme of hypergeometric orthogonal polynomials and satisfy discrete orthogonality relations~\cite{Koekoek}. 

The Racah polynomials $R_x(k; \alpha, \beta, \gamma, \delta)$ are defined in terms of the terminating hypergeometric series,
\begin{equation}\label{poly-Racah}
R_x(k; \alpha, \beta, \gamma, \delta) =  {}_4F_3 \left( \begin{array}{c} -x, x+\alpha+\beta+1, -k, k+\gamma+\delta+1 \\ \alpha+1, \beta+\delta+1, \gamma+1 \end{array} ; 1 \right).
\end{equation}

These polynomials can be connected to representation theory \cite{zhedanov1988nature,zhedanov1991hidden}, combinatorics \cite{leonard1984parameters,leonard1982orthogonal}, and mathematical physics. They serve as finite-dimensional analogs of the Wilson polynomials and are useful in applications such as Racah coefficients (or $6j$ symbols) in quantum mechanics \cite{wilson1978hypergeometric, racah1942theory}.
The Racah polynomials contain several families as limiting cases, including:
discrete Hahn polynomials, Meixner and Krawtchouk polynomials
  or Laguerre and Hermite polynomials.

For the scope of this paper, we will use the fact that they satisfy a three-term recurrence relation to study in detail the inhomogeneous systems. More specifically, for some suitably chosen parameters in the tridiagonal matrix $ \mathcal{K}$, the spectrum can be solved analytically in terms of orthogonal polynomials.

Below we will consider two cases of orthogonal polynomials: the Krawtchouk polynomials, and the dual Hahn polynomials, but obviously the techniques apply to the full scheme.

\subsection{Potential related to the Krawtchouk polynomials}

As a first example of the general strategy described in Sec.~\ref{sec:inhMod}, we consider an inhomogeneous model based on the Krawtchouk polynomials. We follow the approach developed in \cite{Crampe:2019upj} for fermionic models and adapt it to the bosonic Hamiltonian \eqref{inhomogHam}.

\begin{figure}
    \centering
    \includegraphics[width=.45\linewidth]{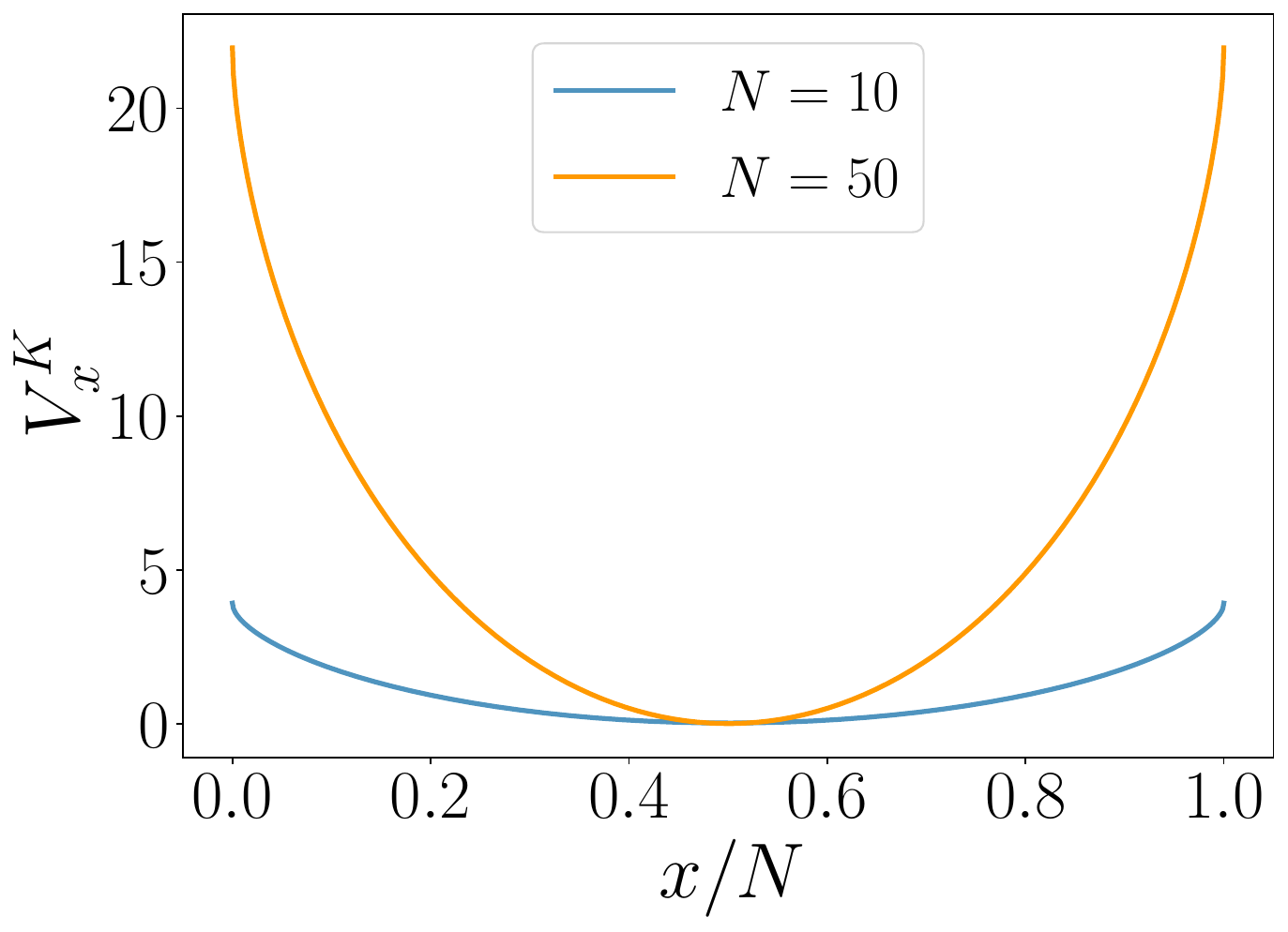}\hspace{.5cm}
     \includegraphics[width=.45\linewidth]{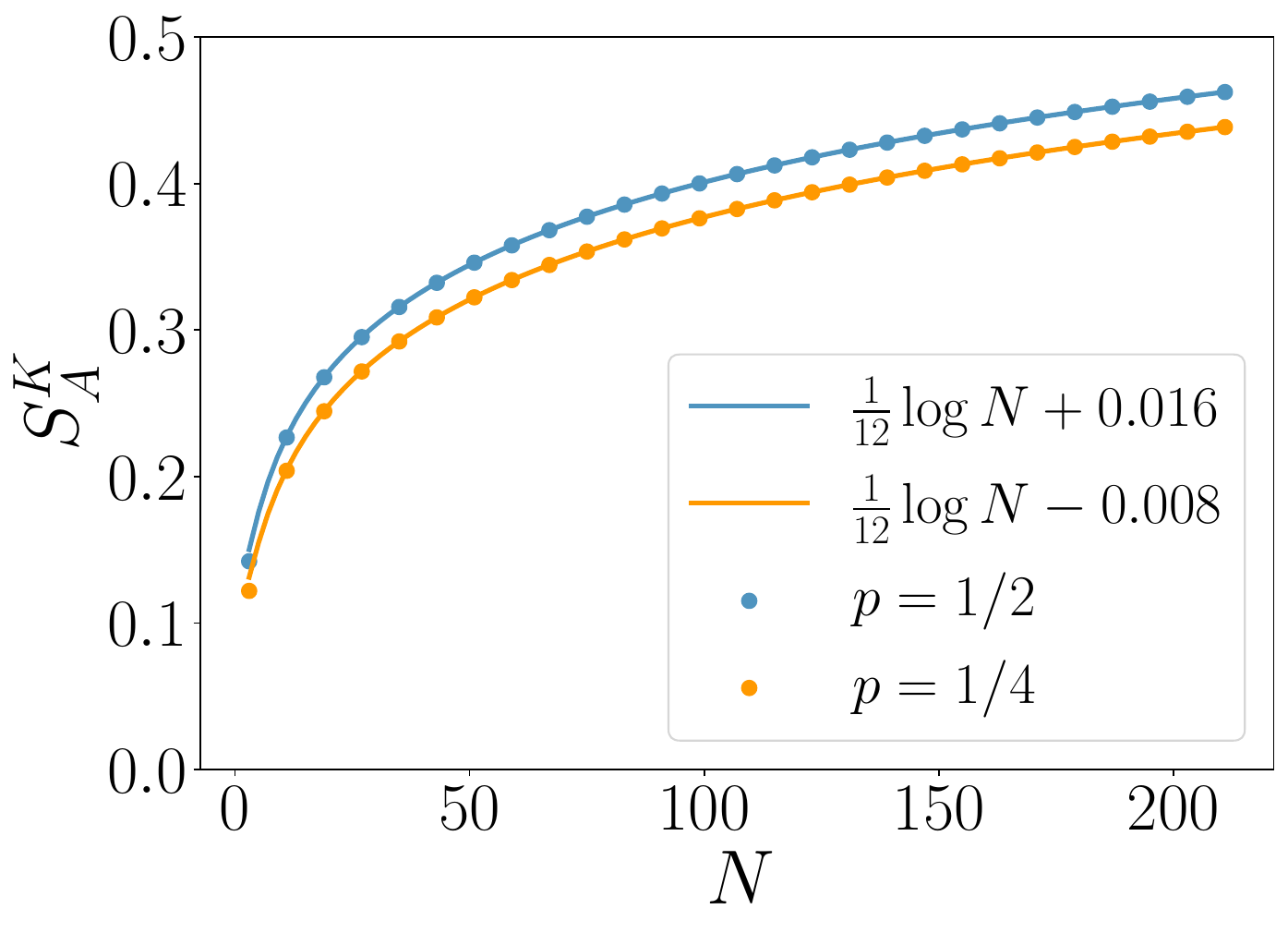}
    \caption{\textit{Left:} Potential $V^K_x$ in Eq. \eqref{eq:VKraw} corresponding to the Krawtchouk chain with $p=1/2$ and different system size~$N$. The effective massless region appears to shrink for increasing values $N$. This could look contradictory to Fig.~\ref{fig:VxGen}, but it is simply an artifact of 
    plotting the potential as a function of $x/N$ instead of $x$. Indeed, the effective massless region scales as $N^{1/2}$, and we represent it as a function of~$x/N$. \textit{Right:} Entanglement entropy of the Krawtchouk chain $S_A^K$ as a function of $N$ where $A=\{0,1,\dots,x_0=pN\}$ for different values of $p$. The symbols are obtained by exact diagonalization and the solid lines are given by Eq. \eqref{eq:SKraw} with constants which are obtained by numerical fit.}
   \label{fig:VKSK}
\end{figure}

Krawtchouk polynomials correspond to the limit $\gamma+1=-N$, $\delta=t^2$, $\alpha=pt$, $\beta=(1-p)t$ with $t\to\infty$ in Eq.~\eqref{poly-Racah}. They read
\begin{equation}
 K_x(k,p;N)=   {}_2F_1 \left( \begin{array}{c} -x,  -k \\ -N  \end{array} ; \frac1p \right) \,,
\end{equation}
where $0<p<1$ is a parameter. 
In terms of the coupling constants of the Hamiltonian  \eqref{inhomogHam}, the Krawtchouk case corresponds to the parameters \cite{Crampe:2019upj}
\begin{equation}
    J_x = \sqrt{(N-x)(x+1)p(1-p)}, \qquad
    B_x = Np + x(1-2p) + m_K^2 \,,
\end{equation}
where $B_x$ is shifted by a constant mass term $m_K\geqslant 0$ compared to Ref.~\cite{Crampe:2019upj} in order to ensure a positive potential. The normalized eigenvectors of $ \mathcal{K}$ in Eq. \eqref{Kinhomog} are not affected by the mass term, and read
\begin{equation}
    u_x(\Lambda_k) = (-1)^x \sqrt{\Big(\frac{p}{1-p}\Big)^{x+k}(1-p)^N  \binom{N}{x} \binom{N}{k}}K_x(k,p,N),
\end{equation}
while the eigenvalues are shifted,
\begin{equation}
   \Lambda_k = k+m_K^2, \quad k=0,1,\dots, N.
\end{equation}
The model describes bosons with inhomogeneous couplings $J_x$ in an inhomogeneous potential\footnote{The potential $V^K_x$ and the entanglement entropy $S^K_A$ are written with a superscript $K$ to highlight the fact that they pertain to the Krawtchouk case. }
\begin{equation}\label{eq:VKraw}
    V^K_x = Np + x(1-2p) + m_K^2-  \sqrt{(N-x)(x+1)p(1-p)} -  \sqrt{(N-x+1)x p(1-p)}.
\end{equation}
The potential is represented in the left panel of Fig. \ref{fig:VKSK} for different values of $N$. For large $N$,
it has a minimum at $x_0=Np$ where it reads $V^K_{x_0} = (m_K^2-\frac 12) + \mathcal{O}(N^{-1})$. We thus choose $m_K=\frac{1}{\sqrt{2}}$ such that the Krawtchouk potential vanishes at $x_0=Np$ in the large-$N$ limit. Hence, in this limit the theory is locally massless around $x_0$. Following the discussion in Sec. \ref{sec:potential}, we expand the potential around the minimum to find the effective regions where the theory is massless. We find\footnote{This computation can be done conveniently with {\tt Mathematica}, using the built-in function {\tt Series[]} and explicit values of $a$.}
\begin{equation}
    V^K_{x_0 \pm y N^a} = \frac{y^2}{4p(1-p)}N^{2a-1} + \mathcal{O}(N^{\max(a-1, \ 3a-2)}). 
\end{equation}
The potential vanishes in the large-$N$ limit in the region $x_0 \pm y N^a$ for $a \in [0,\frac 12)$. This corresponds to $a^*=\frac 12$,
indicating that the effective massless region scales as $\N_{\rm eff} \sim N^{1/2}$. We thus expect the entanglement entropy to scale as
\begin{equation}\label{eq:SKraw}
    S^K_{A} \sim \frac{1}{12} \log N
\end{equation}
 for $A=\{0,1,\dots,x_0=pN\}$ and $m_K=\frac{1}{\sqrt{2}}$. This scaling thus corresponds to an effective central charge $c_{\rm eff}=1/2$. 

\begin{figure}
    \centering
  \includegraphics[width=.45\linewidth]{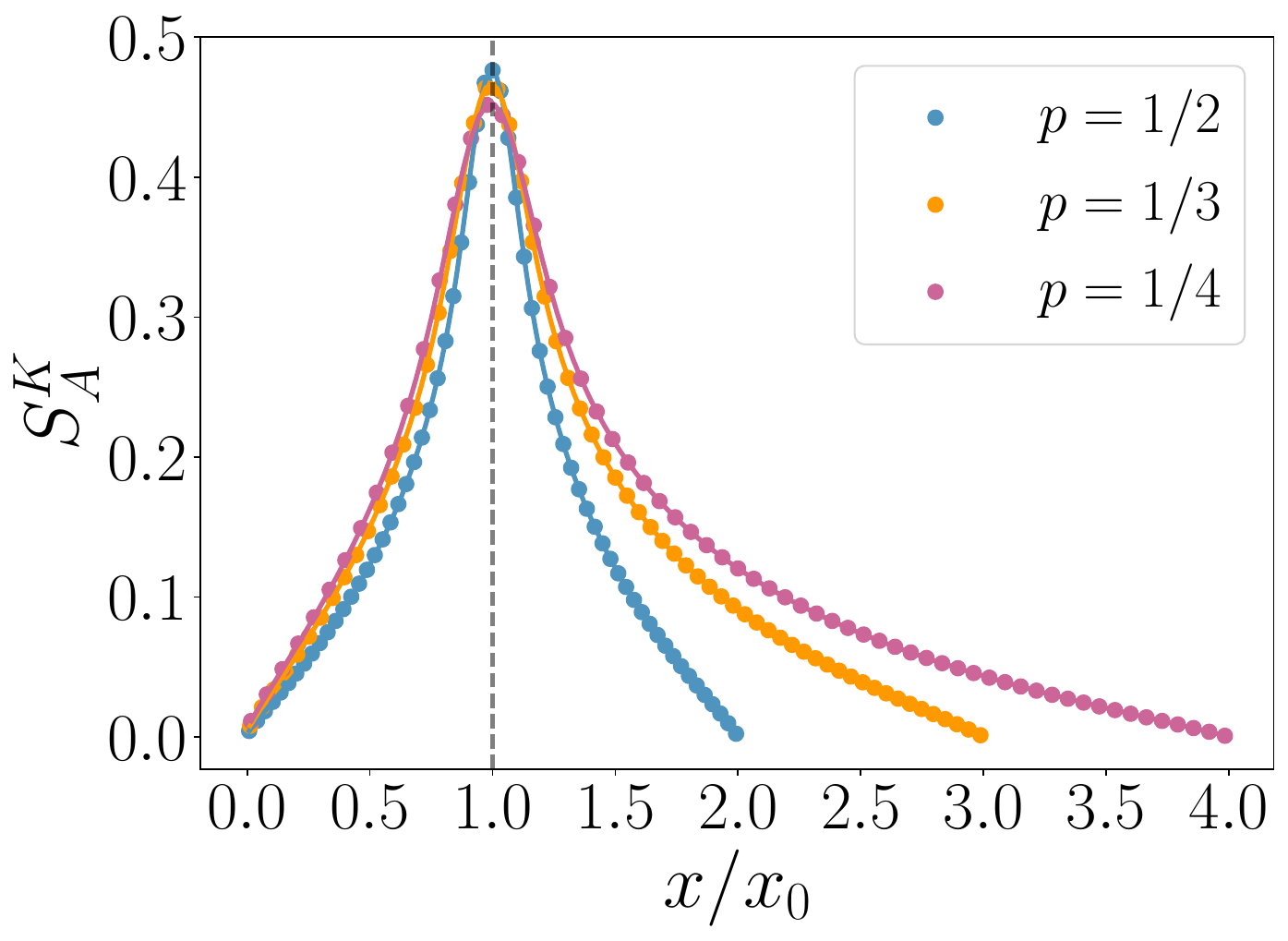} \hspace{.5cm}
   \includegraphics[width=.45\linewidth]{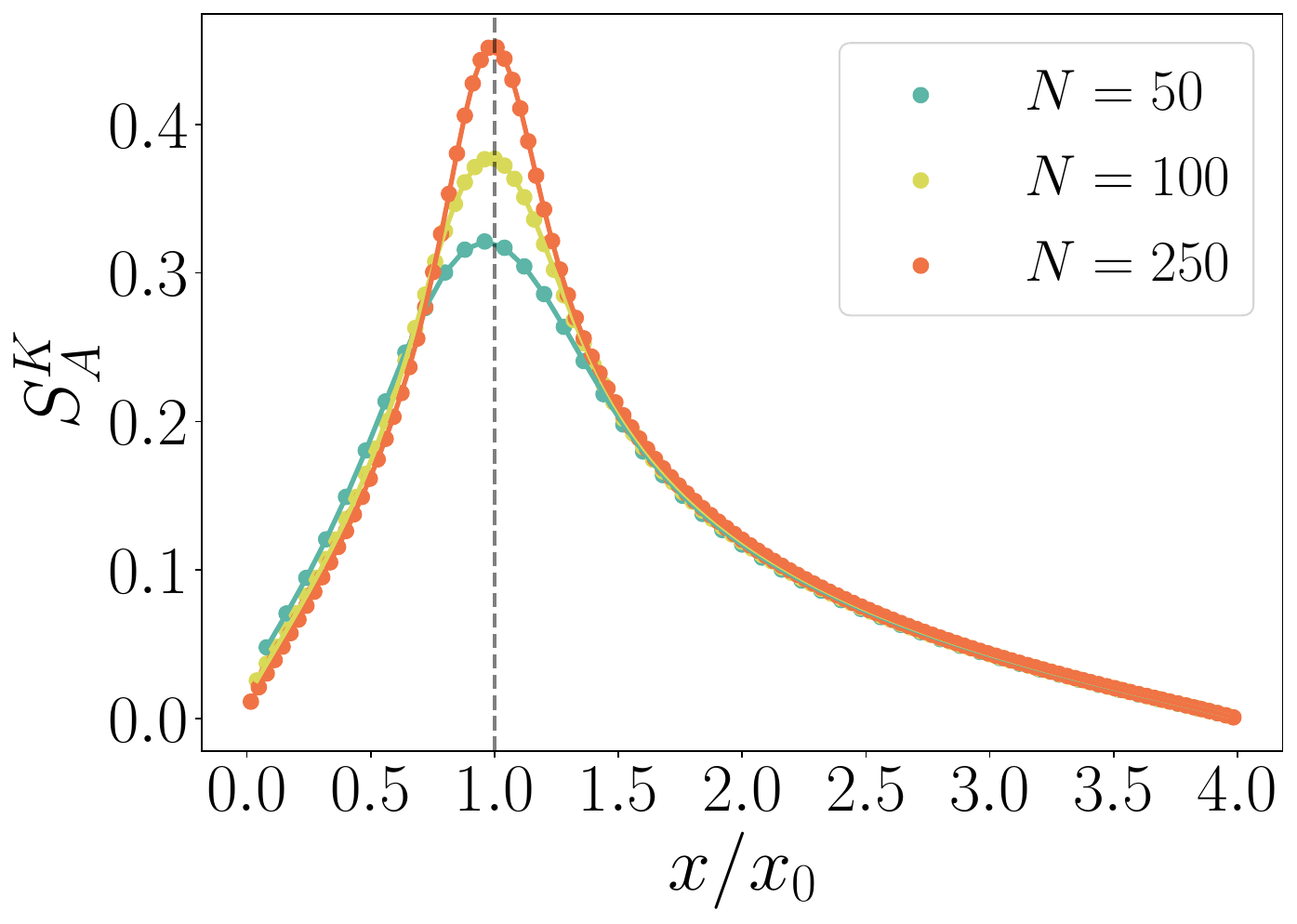}
    \caption{\textit{Left:} Entanglement entropy for fixed $N=250$ as a function of $x/x_0$ for different values of $p$. The entanglement entropy has a sharp peak around $x=x_0$, as a expected. \textit{Right:} Entanglement entropy for fixed $p=1/4$ as a function of $x/x_0$ for increasing values of $N$. In the effective massive region the entropy is a constant with respect to $N$, whereas it grows in the massless region centered on $x_0$. On both panels, the subsystem considered is $A=\{0,1,\dots,x\}$, the symbols are obtained by exact diagonalization, and the solid lines serve as a guide to the eye. }
   \label{fig:S1Kraw}
\end{figure}

 To numerically investigate the entanglement entropy, we use the machinery introduced in Sec.~\ref{sec:EECa}. In particular, we employ Eqs.~\eqref{eq:corrXP}, \eqref{eq:corrC} and \eqref{eq:SA} with the matrices $\cal K$ and $U$ pertaining to the Krawtchouk case.
 We show the comparison between the prediction of Eq.~\eqref{eq:SKraw} and the exact diagonalization in the right panel of Fig.~\ref{fig:VKSK}, and find excellent agreement. As discussed in Sec.~\ref{sec:potential}, the same conclusion holds for a subsystem $A=\{0,1,\dots,x\}$ with $|x-x_0|<\mathcal{O}(\N_{\rm eff})$. To analyze this behavior further, we represent the entanglement entropy as a function of boundary point $x$ for a large value of $N$ in the left panel of Fig.~\ref{fig:S1Kraw}. We observe  peak of width $\sim \N_{\rm eff}$ near the critical point~$x_0$, reflecting a region where the entropy scales as $ \mathcal{O}(\log N)$ instead of $ \mathcal{O}(1)$. Finally, in the right panel of Fig.~\ref{fig:S1Kraw}, we again represent the entanglement entropy as a function of $x$, but where each curve corresponds to a different value of~$N$. Far from $x_0$, all the curves collapse, indicating that indeed the entanglement entropy is a constant with respect to $N$, whereas in the vicinity of $x_0$ the entropy grows with $N$, as expected.

\subsection{Potential related to the dual Hahn polynomials}
The second example of our general strategy is based on the dual Hahn polynomials.
The dual Hahn polynomials correspond to the case $\alpha+1=N$ and $\beta\to\infty$ in \eqref{poly-Racah}, leading to
\begin{equation}
H_x(k,\gamma,\delta;N)={}_3F_2 \left( \begin{array}{c} -x, -k, k+\gamma+\delta+1 \\ -N, \gamma+1 \end{array} ; 1 \right) \,,
\end{equation}
where $\gamma$ and $\delta$ are parameters $0<\gamma \leqslant \delta$.

\begin{figure}
    \centering
  \includegraphics[width=.45\linewidth]{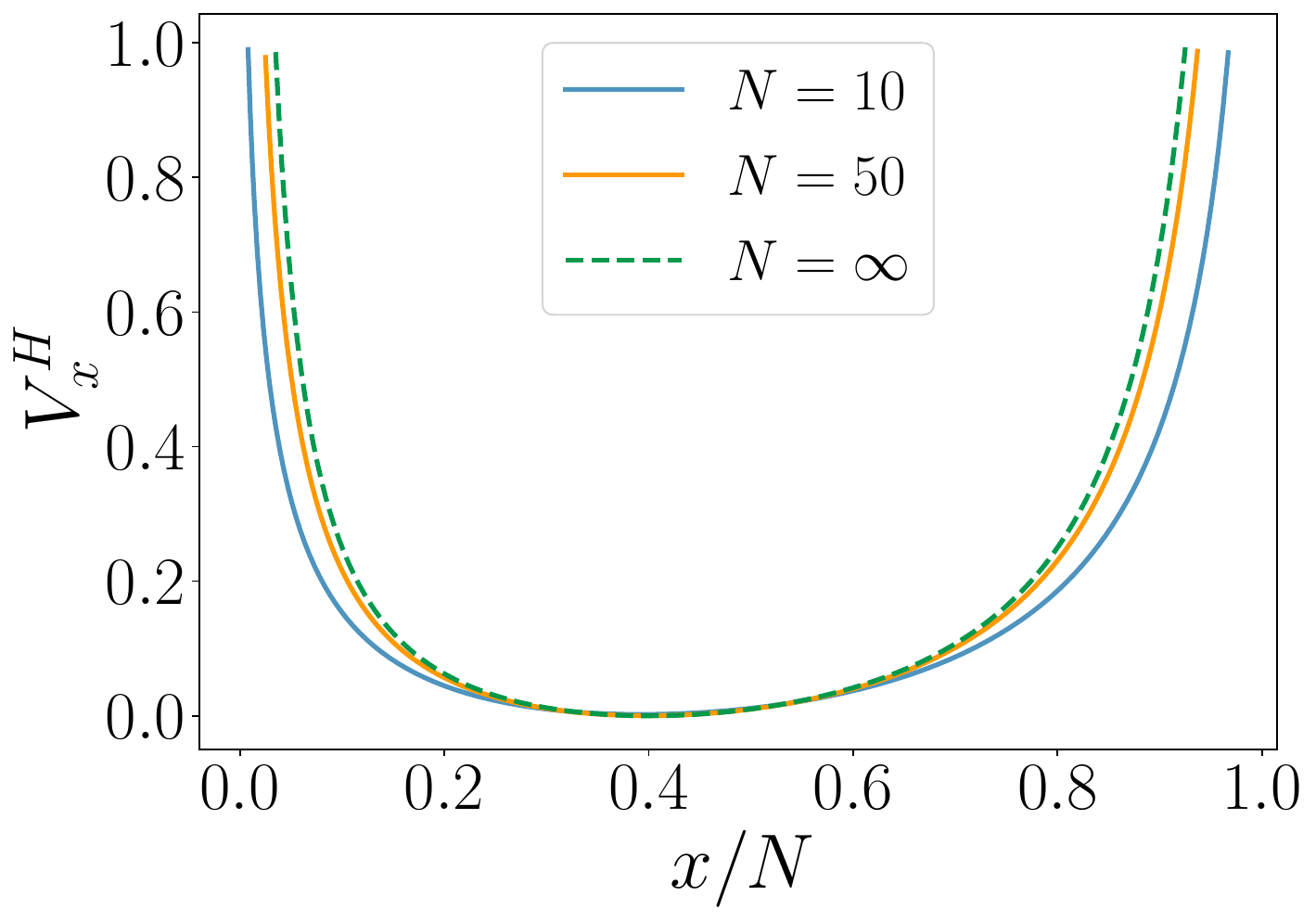}\hspace{.5cm}
  \includegraphics[width=.45\linewidth]{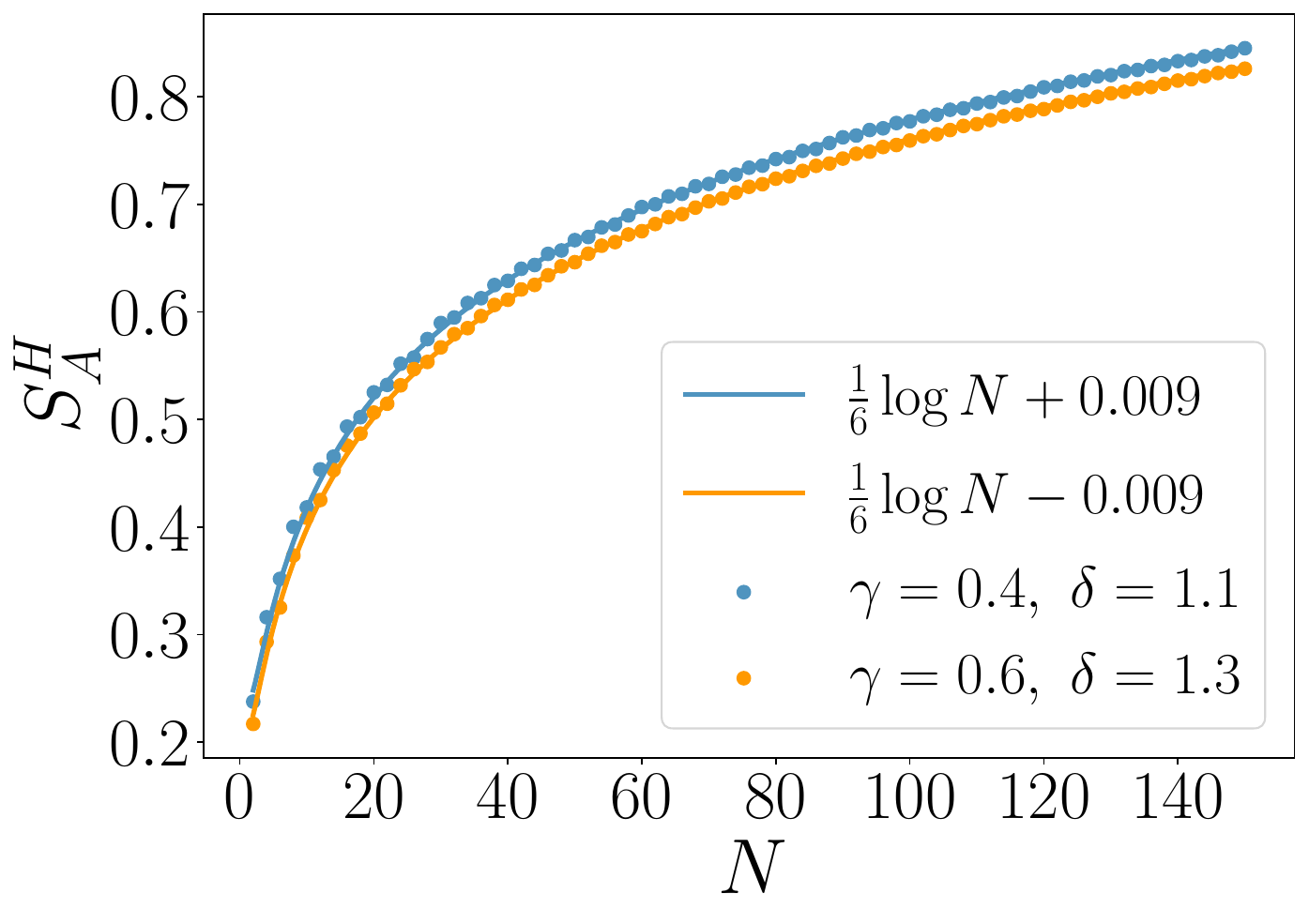}
    \caption{\textit{Left:} Potential $V^H_x$ in Eq. \eqref{eq:VdH} corresponding to the dual Hahn chain with $\gamma=2/5$, $\delta=3/5$ and different values of~$N$. Since the massless region scales as $N$, it has a constant size as a function of $x/N$, unlike the Krawtchouk case, see Fig.~\ref{fig:VKSK}. \textit{Right:} Entanglement entropy of the dual Hahn chain $S_A^H$ as a function of $N$ with $A=\{0,1,\dots,x_0=N\gamma/(\gamma+\delta)\}$ for different values of $\gamma,\delta$. The symbols are obtained by exact diagonalization and the solid lines are given by Eq. \eqref{eq:SdH} with constants that are obtained by numerical fit. }
   \label{fig:VHSH}
\end{figure}

\begin{figure}
    \centering
   \includegraphics[width=.45\linewidth]{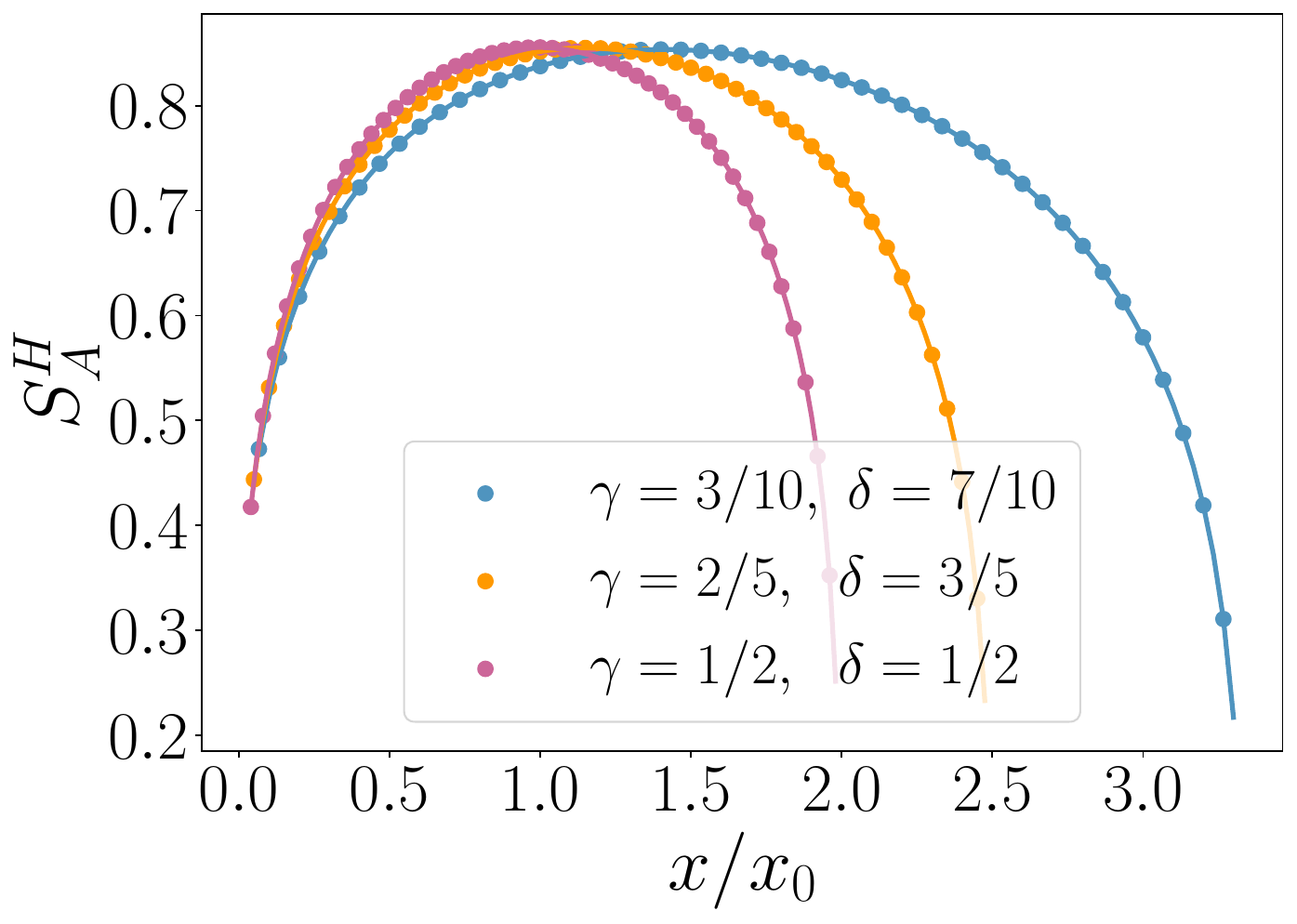}\hspace{.5cm}
   \includegraphics[width=.45\linewidth]{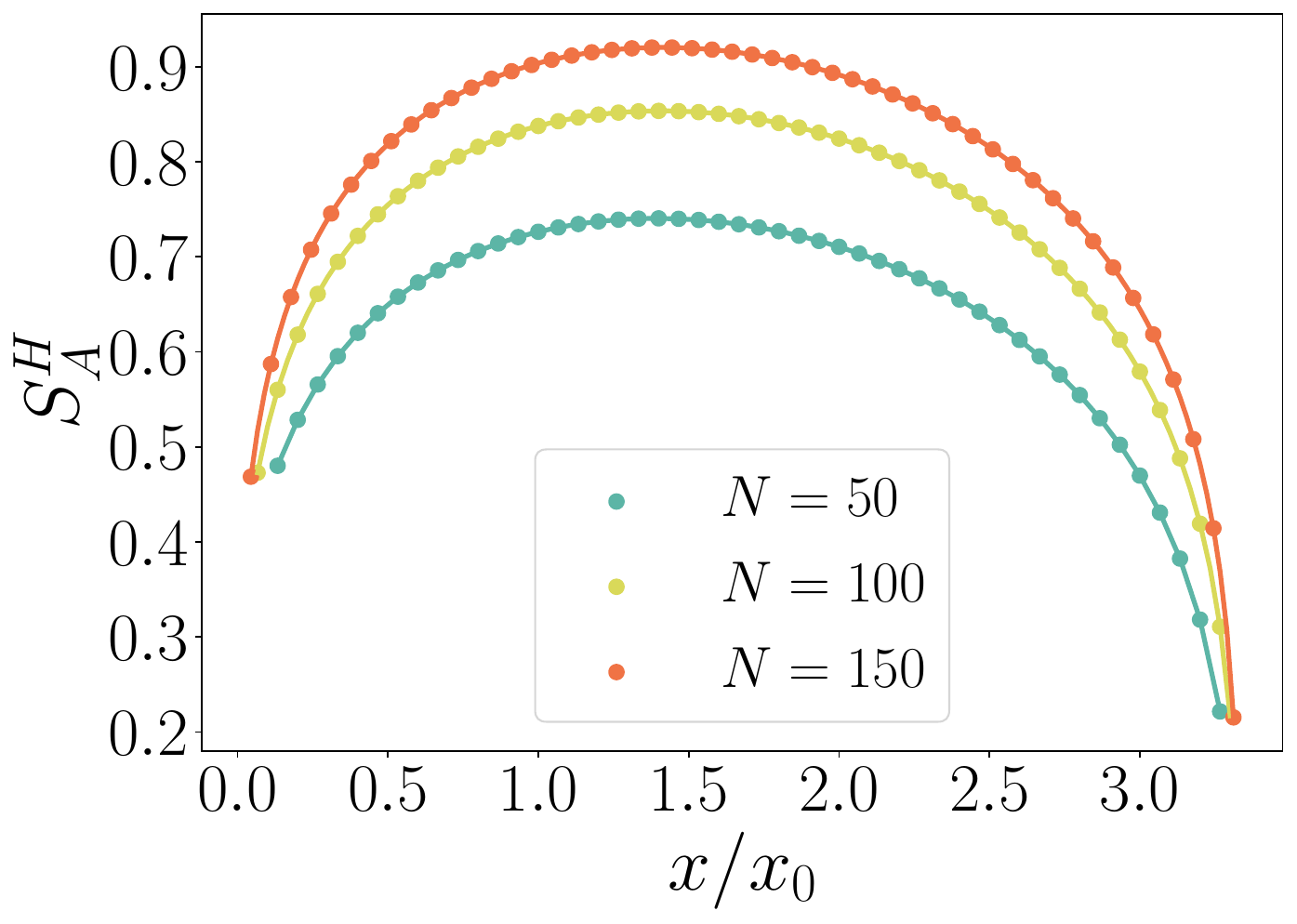}
    \caption{\textit{Left:} Entanglement entropy for fixed $N=100$ as a function of $x/x_0$ for different values of $\gamma,\delta$. As opposed to the Krawtchouk case, the entanglement entropy does not have a sharp peak. \textit{Right:} Entanglement entropy for fixed $\gamma=3/10$, $\delta=7/10$ as a function of $x/x_0$ for increasing values of $N$. On both panels, the subsystem considered is $A=\{0,1,\dots,x\}$, the symbols are obtained by exact diagonalization, and the solid lines serve as a guide to the eye.}
   \label{fig:S1dH}
\end{figure}

The parameters in the matrix $ \mathcal{K}$ in \eqref{Kinhomog} read in this case \cite{Crampe:2019upj}
\begin{equation}
    J_x = \sqrt{(x+1)(x+\gamma+1)(N-x)(N-x+\delta)}, \qquad
    B_x = N+(N-x)(2x+\gamma)+ x\delta+m^2_H.
\end{equation}
The eigenvectors are given by
\begin{equation}
u_x(\Lambda_k) =\sqrt{ 
{N \choose k}{\gamma+x\choose x}{\delta+N-x\choose N-x}
\frac{(2k+\gamma+\delta+1)(\gamma+1)_{k} N!}
	{(k+\gamma+\delta+1)_{N+1} (\delta+1)_{k}} 
 }\,
H_{x}(k,\gamma,\delta;N)
\end{equation}
with eigenvalues
\begin{equation}
\Lambda_k = k(k+\gamma+\delta+1)+m_H^2, \quad k=0,1,\dots, N.
\end{equation}
From Eq. \eqref{eq:Vx}, the potential reads\footnote{The potential $V^H_x$ and the entanglement entropy $S^H_A$ are written with a superscript $H$ to highlight the fact that they pertain to the dual Hahn case. }
\begin{equation}\label{eq:VdH}
    V^{H}_x = N+(N-x)(2x+\gamma)+ x\delta+m_H^2 - \sqrt{(x+1)(x+\gamma+1)(N-x)(N-x+\delta)}- \sqrt{x(x+\gamma)(N-x+1)(N-x+1+\delta)}.
\end{equation}
We represent this potential in the left panel of Fig.~\ref{fig:VHSH}. To proceed, we follow the same steps as for the Krawtchouk case. First, in the large-$N$ limit, the potential is minimal for $x_0=N \gamma/(\gamma+\delta)$, and reads $V^{H}_{x_0} = (m_H^2-\frac{\gamma+\delta}{2})+ \mathcal{O}(N^{-1})$. In order to have a region where the potential is locally zero, and hence a locally massless theory, we choose $m_H = \sqrt{\frac{\gamma+\delta}{2}}$. Expanding the potential around the minimum, we find 
\begin{equation}
    V^{H}_{x_0 \pm y N^a} = \frac{y^2(\gamma+\delta)^4}{4\gamma \delta}N^{2a-2} + \mathcal{O}(N^{\max(a-2, \ 3a-3)}), 
\end{equation}
which vanishes for $a \in [0,1)$. This corresponds to $a^*=1$, and hence $\N_{\rm eff}\sim N$. We thus expect the entanglement entropy to scale as
\begin{equation}\label{eq:SdH}
    S^H_{A}\sim \frac 16 \log N
\end{equation}
for $A=\{0,1,\dots,x_0=N \gamma/(\gamma+\delta)\}$ and $m_H = \sqrt{\frac{\gamma+\delta}{2}}$, corresponding to the effective central charge $c_{\rm eff}=1$. To investigate the entanglement entropy numerically, we use the results of Sec.~\ref{sec:EECa} and adapt them to the dual Hahn case. We compare the  prediction of Eq.~\eqref{eq:SdH} with exact numerical results in the right panel of Fig. \ref{fig:VHSH}, and find very good agreement. As for the Krawtchouk case, we further investigate the behavior of the entanglement entropy as a function of $x$ when the subsystem is $A=\{0,1,\dots,x\}$ in both panels of Fig.~\ref{fig:S1dH}. In the left panel, we observe that the entropy does not have a sharp peak, in agreement with the fact that the size of the effective massless region scales as the total system size, $\N_{\rm eff}\sim N$. We note that the entropy is not necessarily maximal for $x=x_0$, which is likely due to the fact that $x_0$ is not necessarily in the center of the massless region. Indeed, on the left panel of Fig.~\ref{fig:VHSH} we clearly see that the dual Hahn potential is not symmetric. Finally, in the right panel of Fig.~\ref{fig:S1dH}, we observe that the entanglement entropy grows with $N$ in essentially the whole system, in stark contrast with the Krawtchouk case. This behavior is once again in agreement with our general interpretation of scaling the size of the effective massless region. 

\subsection{An infinite family of effective central charges}
As a final example of solvable lattice model based on orthogonal polynomials, we interpolate between the dual Hahn and the Krawtchouk cases.
In order to motivate the interpolating model, we observe that
the Krawtchouk polynomials can be obtained through the limit $\gamma=pt$, $\delta=(1-p)t$ with $t\to\infty$ in dual Hahn polynomials. In this limit (with $N$ fixed), the two potentials satisfy
\begin{equation}\label{eq:VHtVx}
    \lim_{t\to \infty}\frac{V^H_x}{t} = V^K_x.
\end{equation}
The normalization in the LHS of Eq. \eqref{eq:VHtVx} ensures that the mass is finite for all values of $t$. Indeed, the associated mass term is $m_H^2/t = 1/2=m_K^2$.

\begin{figure}
    \centering
    \includegraphics[width=.85\linewidth]{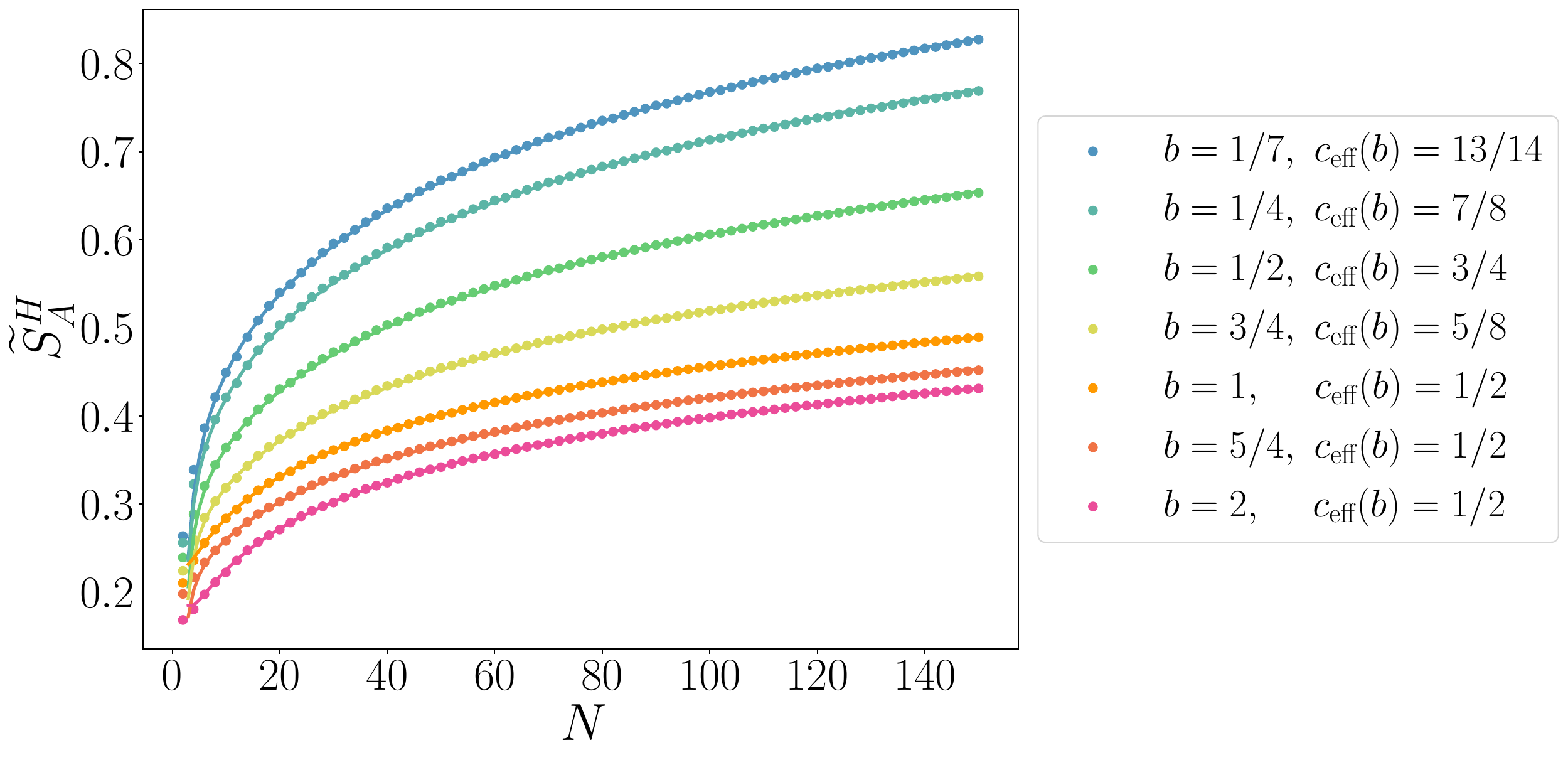}

    \caption{Entanglement entropy of the dual Hahn chain with parameters $\gamma=p N^b$ and $\delta=(1-p)N^b$ as a function of $N$ with $p=0.4$ and different values of $b$. The symbols are obtained by exact diagonalization and the solid lines are obtained from Eq.~\eqref{eq:SAdHtoK} with additional constants and subleading terms obtained from numerical fit.}
   \label{fig:S1dHahn}
\end{figure}

Since in our investigations of the entanglement scaling we also consider the limit $N\to \infty$, a natural generalization of the above limit is to replace $t$ by a power of~$N$, and to   
consider the modified dual Hahn potential
\begin{equation}
    \widetilde{V}^H_x=V^H_x N^{-b}
\end{equation}
with 
\begin{equation}
    \gamma=pN^b\,,\quad \delta=(1-p)N^b\,,\quad b>0.
\end{equation}
The potential $\widetilde{V}^H_x$ is minimal at $x_0=pN$. However, depending on the value of $b$, the factors $(N-x+\delta)$ and $(N-x+1+\delta)$ in $\widetilde{V}^H_x$ behave differently, leading to different values of $\N_{\rm eff}$. We obtain the scaling
\begin{equation}
    \widetilde{V}^H_{x_0 \pm y N^a} = \frac{y^2}{4p(1-p)}\times \left\{
\begin{array}{lll}
    N^{2a - 2+b} &+ \mathcal{O}(N^{\max(a-2+b, \ 3a-3+b, \ 2a+2b-3)}), & 0\leqslant  b  < 1, \\[5pt]
    N^{2a - 1}   &+ \mathcal{O}(N^{\max(a-1,\ 3a-2)}),   & b\geqslant1, 
\end{array}
\right. 
\end{equation}
which corresponds to
\begin{equation}
    \N_{\rm eff}\sim  \left\{
\begin{array}{ll}
   N^{\frac{2-b}{2}},  & 0\leqslant b<1 , \\[5pt]
    N^{\frac{1}{2}} ,  & b\geqslant1.
\end{array}
\right. 
\end{equation}
We thus expect the entanglement entropy of the interval $A=\{0,1,\dots,x_0\}$ to scale as 
\begin{equation}\label{eq:SAdHtoK}
    \widetilde{S}^H_{A} \sim \frac{c_{\rm eff}(b)}{6} \log N, \qquad   c_{\rm eff}(b) =  \left\{
\begin{array}{ll}
   \frac{2-b}{2}, & 0\leqslant b <1,  \\[5pt]
    \frac{1}{2},   & b\geqslant1,
\end{array}
\right. 
\end{equation}
leading to a continuum of 
effective central charge $c_{\rm eff}(b)$. In particular, for $b=\frac{12}{q(q+1)}$ and integer $q>2$, it encompasses the central charges associated to unitary minimal models, whereas for rational $q$ one gets non-unitary minimal models central charges \cite{francesco2012conformal}. We compare the scaling prediction of Eq.~\eqref{eq:SAdHtoK} with exact numerical results for various values of~$b$ in Fig. \ref{fig:S1dHahn}, and find very good agreement. 

\section{Conclusion}\label{sec:ccl}

In this paper, we investigated the ground-state entanglement entropy in inhomogeneous free-boson models in one spatial dimension. We developed a powerful method to extract the leading term in the entanglement scaling, based on the analytic properties of the inhomogeneous potential. The key idea is to identify regions where the theory is massless, and to characterize the scaling of their effective size $\N_{\rm eff}\sim N^{a^*}$ in the large-$N$ limit. For regions whose contact point lies in the massless region, the entanglement diverges logarithmically with an effective central charge~$c_{\rm eff}=a^*$.
In contrast, for subsystems whose contact point is in a massive region, the entanglement entropy scales as a constant with respect to the system size $N$. This method applies to arbitrary inhomogeneous models with smooth potential, and we illustrated it for a family of exactly-solvable models based on orthogonal polynomials of the Askey scheme, finding a perfect match with the exact numerical results. 

There are several points of interest for future investigations. First, it would be important to apply our method to other inhomogeneous bosonic systems. For instance, one could consider more general potentials, where no analytical calculations can be performed, or other solvable models not related to orthogonal polynomials. Second, one could study the impact of the inhomogeneous potential on other entanglement measures, such as the logarithmic negativity, to further probe the effect of inhomogeneities on the entanglement structure of quantum many-body states. Generalizing our results in the case of higher-dimensional systems is also promising. 
For inhomogeneous fermionic systems, the curved-space CFT approach is a very powerful method to extract the entanglement scaling in the thermodynamic limit~\cite{DSVC17,Finkel:2021gji,Finkel:2020lgf}. A natural question would thus be to generalize the approach and apply it to inhomogeneous bosonic systems in order to refine the results we obtained in this paper, and in particular access the constant and subleading terms. However, this appears to be non trivial. Technically, the difficulty arises from the fact that the inhomogeneity coefficients $J_x,V_x$ in the Hamiltonian \eqref{inhomogHam} only pertain to the fields $\phi_x$, whereas the terms involving $\pi_x$ are homogeneous. In the continuous limit, this difference in the two fields generates consistency relations that can only be satisfied for $J_x=1$, i.e., the homogeneous case with an inhomogeneous potential. We leave this important point for future investigations. 

\section*{Acknowledgments}

GP thanks Cl\'ement Berthiere for useful comments on the manuscript. PAB holds an Alexander-Graham-Bell scholarship from the Natural Sciences and Engineering Research Council (NSERC) of
Canada. RN is supported in part by the National Science Foundation under grant PHY 2310594, and by a Cooper fellowship. LV is funded in part by a Discovery Grant from NSERC.

\providecommand{\href}[2]{#2}\begingroup\raggedright\endgroup

\end{document}